\documentclass[twocolumn,superscriptaddress,aps,preprintnumbers,amsmath,amssymb,prl,nofootinbib]{revtex4-1}

\usepackage{graphicx}
\usepackage{epstopdf}
\usepackage{dcolumn}
\usepackage{bm}
\usepackage{hyperref}
\usepackage{color}

\begin{document}


\def\a{\alpha}
\def\b{\beta}
\def\c{\varepsilon}
\def\d{\delta}
\def\e{\epsilon}
\def\f{\phi}
\def\g{\gamma}
\def\h{\theta}
\def\k{\kappa}
\def\l{\lambda}
\def\m{\mu}
\def\n{\nu}
\def\p{\psi}
\def\q{\partial}
\def\r{\rho}
\def\s{\sigma}
\def\t{\tau}
\def\u{\upsilon}
\def\v{\varphi}
\def\w{\omega}
\def\x{\xi}
\def\y{\eta}
\def\z{\zeta}
\def\D{\Delta}
\def\G{\Gamma}
\def\H{\Theta}
\def\L{\Lambda}
\def\F{\Phi}
\def\P{\Psi}
\def\S{\Sigma}

\def\o{\over}
\def\beq{\begin{eqnarray}}
\def\eeq{\end{eqnarray}}
\newcommand{\gsim}{ \mathop{}_{\textstyle \sim}^{\textstyle >} }
\newcommand{\lsim}{ \mathop{}_{\textstyle \sim}^{\textstyle <} }
\newcommand{\vev}[1]{ \left\langle {#1} \right\rangle }
\newcommand{\bra}[1]{ \langle {#1} | }
\newcommand{\ket}[1]{ | {#1} \rangle }
\newcommand{\EV}{ {\rm eV} }
\newcommand{\KEV}{ {\rm keV} }
\newcommand{\MEV}{ {\rm MeV} }
\newcommand{\GEV}{ {\rm GeV} }
\newcommand{\TEV}{ {\rm TeV} }
\newcommand{\1}{\mbox{1}\hspace{-0.25em}\mbox{l}}
\newcommand{\headline}[1]{\noindent{\bf #1}}
\def\diag{\mathop{\rm diag}\nolimits}
\def\Spin{\mathop{\rm Spin}}
\def\SO{\mathop{\rm SO}}
\def\O{\mathop{\rm O}}
\def\SU{\mathop{\rm SU}}
\def\U{\mathop{\rm U}}
\def\Sp{\mathop{\rm Sp}}
\def\SL{\mathop{\rm SL}}
\def\tr{\mathop{\rm tr}}
\def\mpl{M_{PL}}

\def\IJMP{Int.~J.~Mod.~Phys. }
\def\MPL{Mod.~Phys.~Lett. }
\def\NP{Nucl.~Phys. }
\def\PL{Phys.~Lett. }
\def\PR{Phys.~Rev. }
\def\PRL{Phys.~Rev.~Lett. }
\def\PTP{Prog.~Theor.~Phys. }
\def\ZP{Z.~Phys. }

\def\dd{\mathrm{d}}
\def\ff{\mathrm{f}}
\def\BH{{\rm BH}}
\def\inf{{\rm inf}}
\def\ev{{\rm evap}}
\def\eq{{\rm eq}}
\def\SM{{\rm sm}}
\def\Mpl{M_{\rm Pl}}
\def\GeV{{\rm GeV}}
\newcommand{\Red}[1]{\textcolor{red}{#1}}


\title{Revisiting the Minimal Chaotic Inflation Model}

\author{Keisuke Harigaya}
\affiliation{ICRR, University of Tokyo, Kashiwa, Chiba 277-8582, Japan}
\author{Masahiro Ibe}
\affiliation{ICRR, University of Tokyo, Kashiwa, Chiba 277-8582, Japan}
\affiliation{Kavli IPMU (WPI), UTIAS, University of Tokyo, Kashiwa, Chiba 277-8583, Japan}
\author{Masahiro Kawasaki}
\affiliation{ICRR, University of Tokyo, Kashiwa, Chiba 277-8582, Japan}
\affiliation{Kavli IPMU (WPI), UTIAS, University of Tokyo, Kashiwa, Chiba 277-8583, Japan}
\author{Tsutomu T. Yanagida}
\affiliation{Kavli IPMU (WPI), UTIAS, University of Tokyo, Kashiwa, Chiba 277-8583, Japan}
\begin{abstract}
We point out that the prediction of the minimal chaotic inflation model is
 altered if a scalar field takes a large field value close to the Planck scale during inflation
due to a negative Hubble induced mass.
In particular, we show that the inflaton potential is effectively suppressed
at a large inflaton field value in the presence of such a scalar field.
The scalar field may be identified with the standard model Higgs field or flat directions in supersymmetric theory.
With such spontaneous suppression, we find that the minimal chaotic inflation model, 
especially the model with a quadratic potential, is consistent with recent observations 
of the cosmic microwave background fluctuation without modifying the inflation model itself.
\end{abstract}

\date{\today}
\maketitle
\preprint{IPMU 15-0086}

%
Cosmic inflation is the most important paradigm of modern cosmology.
The flatness and the homogeneity of the universe are explained by a quasi-exponential expansion of spacetime in the very early universe~\cite{Guth:1980zm,Kazanas:1980tx}.
Furthermore, so-called slow-roll inflation~\cite{Linde:1981mu,Albrecht:1982wi} (see also~\cite{Starobinsky:1980te})
predicts the almost scale invariant and Gaussian fluctuation of the universe~\cite{Mukhanov:1981xt,Hawking:1982cz,Starobinsky:1982ee,Guth:1982ec,Bardeen:1983qw},
which has been confirmed by observations of
the large scale structure of the universe and the cosmic microwave background (CMB).
Slow-roll inflation is driven by a flat scalar potential of
a scalar field referred as an inflaton.

Among slow-roll inflation models, chaotic inflation~\cite{Linde:1983gd} is the most attractive model,
since it is free from the initial condition problem~\cite{Linde:2005ht}.
Here, let us briefly review the minimal model of chaotic inflation with a quadratic scalar potential~\cite{Linde:1983gd},
\begin{eqnarray}
\label{eq:V1}
V=\frac{1}{2} m^2 \phi^2 \ ,
\end{eqnarray}
where $\phi$ is a real scalar field and $m$ denotes a mass parameter.
In this simplest model, so-called slow-roll conditions are satisfied when 
$\phi$ takes a field value larger than the Planck scale, $M_{\rm PL}\simeq 2.4 \times 10^{18}$\,GeV.
(Hereafter, we occasionally take a unit $M_{\rm PL} = 1$.)
The scalar field $\phi$ plays a role of the inflaton when it starts off from a very large field value 
in the early universe, where the Hubble parameter is given by $H = m\phi/\sqrt{6} $.

From the observed magnitude of the curvature perturbation at the pivot scale, $k_*=0.05$\,Mpc$^{-1}$,
the mass parameter $m$ is determined to be
\begin{eqnarray}
\label{eq:mass}
m\simeq 6\times 10^{-6} \times M_{\rm PL}\ .
\end{eqnarray}
The spectral index of the curvature perturbation $n_s$ and the 
tensor-to-scalar ratio $r$ are, on the other hand, independent of $m$, 
and depend only on the inflaton field value at the corresponding $e$-folding number $N_e$ 
of the pivot scale, $\phi_{N_e}\simeq 2\sqrt{N_e}$;
\begin{eqnarray}
\label{eq:nsr}
n_s &=& 1-\frac{2}{N_e} \simeq 0.967~~(N_e = 60) \ ,\cr
r &=& \frac{8}{N_e} \simeq 0.133~~(N_e = 60) \ .
\end{eqnarray}

The virtue of the choatic inflation model is that it is free from the initial condition problem.
There, 
inflation starts out
with arbitrary or chaotic initial conditions.
It is also advantageous that inflation starts even when the universe is closed at the Planckian time,
since the slow-roll conditions are satisfied for $V\sim M_{\rm PL}^4$ ($\phi \sim M_{\rm PL}^2/m$).
Because of the absence of the initial condition problem and its simplicity,
the chaotic inflation model has attracted great attention
for a long time.

Before applauding the chaotic inflation model, however, we need to address an important question
about the shape of the inflaton potential.
Why is the inflaton potential given by a quadratic term over a Planck scale field value?
Generically, there would be higher dimensional terms of $\phi$ which ruin the flatness of the 
potential.

One of the most attractive ideas addressing the above problem is to use a shift symmetry~\cite{Kawasaki:2000yn}
under which $\phi$ transforms as $\phi+ c$ with $c$ being a real parameter.
The shift symmetry is assumed to be softly broken by a spurion field $m$ which transforms
as $m\to m\phi/(\phi+c)$.
With these assumptions, $\phi$ appears in the scalar potential only through a combination of $m\phi$,
which ensures the dominance of the quadratic term and the flatness of the potential even above the Planck scale field value.

\begin{figure}[tb]
 \begin{center}
  \includegraphics[width=0.9\linewidth]{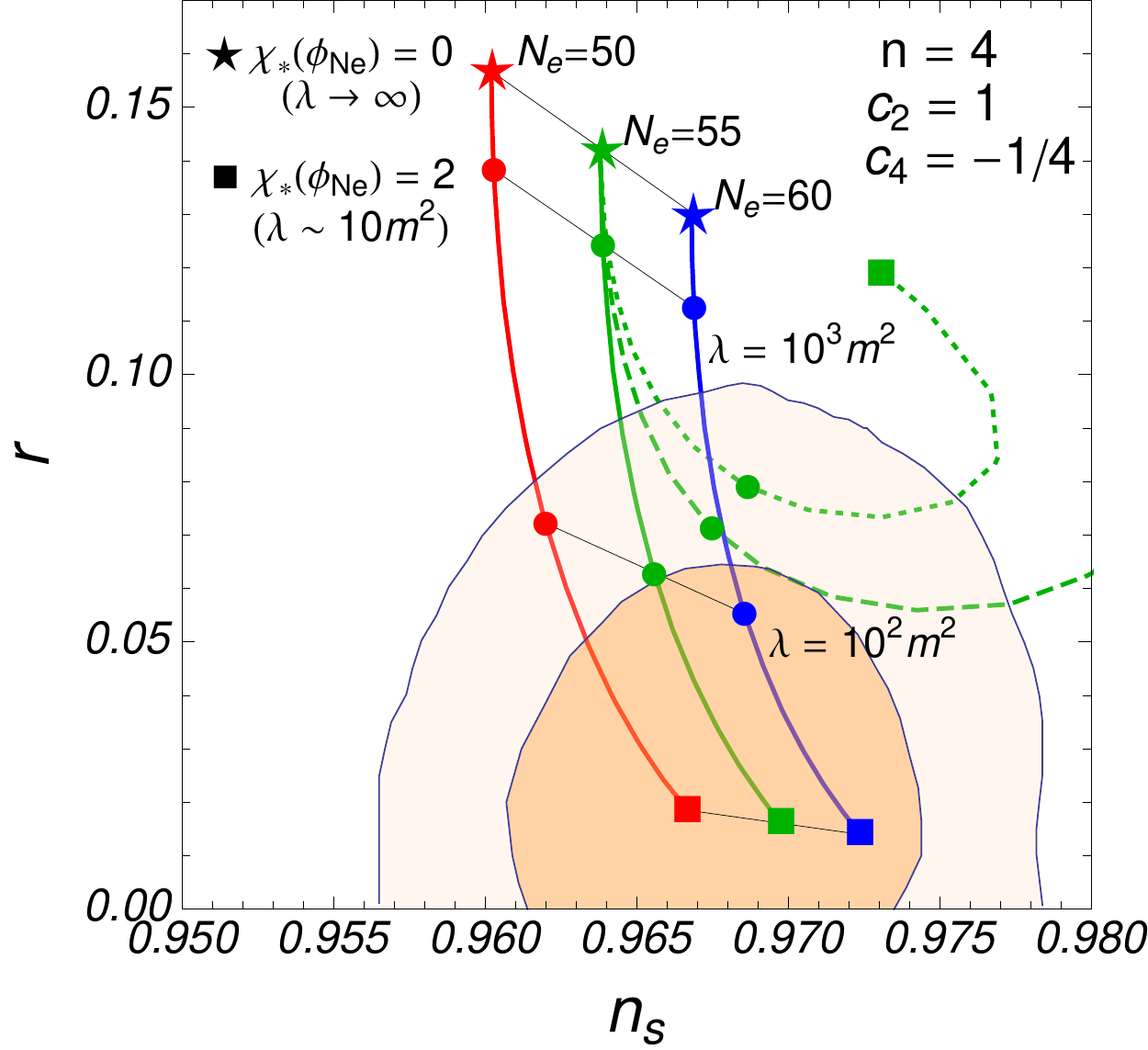}
 \end{center}
\caption{\small\sl 
Predictions on $n_s$ and $r$ of the minimal chaotic 
inflation model with spontaneous suppression. 
The solid lines show the predictions when we change the value of $\lambda$.
There,  $c_2 = 1$, $c_4=-1/4$ and $N_e$ as indicated.
For $N_e = 55$, we also show the predictions for $c_4 = -(1+0.1)/4$ (dashed line) and
$c_4 =- (1+0.2)/4$ (dotted line) for comparison.
The five-points-stars denote the predictions for $\lambda \to \infty$, which 
coincide with the ones without spontaneous suppression.
The points on each line correspond to $\lambda = 10^3m^2$ and $\lambda = 10^2 m^2$ as indicated.
We stop the lines at the filled squares where $\chi_*(\phi_{N_e}) = 2$  ($\lambda \simeq 10 m^2$).
We also show the observational constraints\,\cite{Ade:2015lrj}.
}
\label{fig:nsr}
\end{figure}

\vspace{.5cm}
Despite those successful foundations, however, recent observations of the CMB
seem to disfavor the minimal model with a quadratic inflaton potential~\cite{Ade:2015lrj}
(see also Eq.\,(\ref{eq:nsr}) and Fig.\,\ref{fig:nsr}).
This forces us to modify the minimal model, for example, by introducing further breaking of the shift symmetry
in addition to $m$~\cite{Kallosh:2010ug,Li:2013nfa,Harigaya:2014qza},
or by achieving an inflaton potential with a power law exponent smaller than two~\cite{Silverstein:2008sg,Takahashi:2010ky,Harigaya:2012pg},
or by considering a generic polynomial potential~\cite{Destri:2007pv,Croon:2013ana,Nakayama:2013jka}.

As we will see, however, it is not necessary to modify the inflation model itself to fit the observations.
The inflaton potential is effectively suppressed at a large inflaton field value 
when another scalar field obtains a large value during inflation due to a negative Hubble induced mass.
The scalar field may be identified with the standard model Higgs field or flat directions in supersymmetric theory.
With such  spontaneous suppression of the inflaton potential, we find that the minimal chaotic inflation model 
is consistent with the recent observations.

To illustrate how the suppression occurs, let us introduce a scalar field $\chi$ 
which couples to the inflaton via,
\begin{eqnarray}
V=\frac{1}{2} m^2 \phi^2 \left(1 - c_2 \chi^2  + \cdots\right) + \frac{\lambda}{2n} \chi^{2n}\ .
\label{eq:V2}
\end{eqnarray}
Here, $c_2$ and $\lambda$ are coupling constants, and the ellipses denote
 higher order terms of $\chi$. 
It should be emphasized that the couplings between $\phi$ and  $\chi$ 
do not violate the assumption of  a softly broken shift symmetry.
Rather, it is quite natural for $\chi$ to have such couplings to $\phi$
with ${\cal O}(1)$ coupling constants unless $\chi$ also has a shift symmetry.
As for the scalar potential of $\chi$, we assume a single power low potential of $\chi$ with an exponent $n\ge 2$.
As we will discuss shortly, following arguments can be generalized to a more generic scalar potential of $\chi$.

Now, let us assume that  $c_2$ is positive and $c_2 \gtrsim 1$.
Then, the scalar field $\chi$ obtains a negative induced mass  of ${\cal O}(H^2)$ during inflation,
with which $\chi$ is expelled to a large field value,
\begin{eqnarray}
\label{eq:chistar}
\chi_*(\phi)  \simeq  \left(\frac{c_2}{\lambda}m^2\phi^2\right)^{1/(2n-2)}\ .
\end{eqnarray}
Here, we have also assumed $\lambda > 0$.
Without cancellation, $\chi$ is expected to obtain a mass squared of ${\cal O}(H^2)$ at around $\chi_*(\phi)$.

\begin{figure}[tb]
 \begin{center}
  \includegraphics[width=0.7\linewidth]{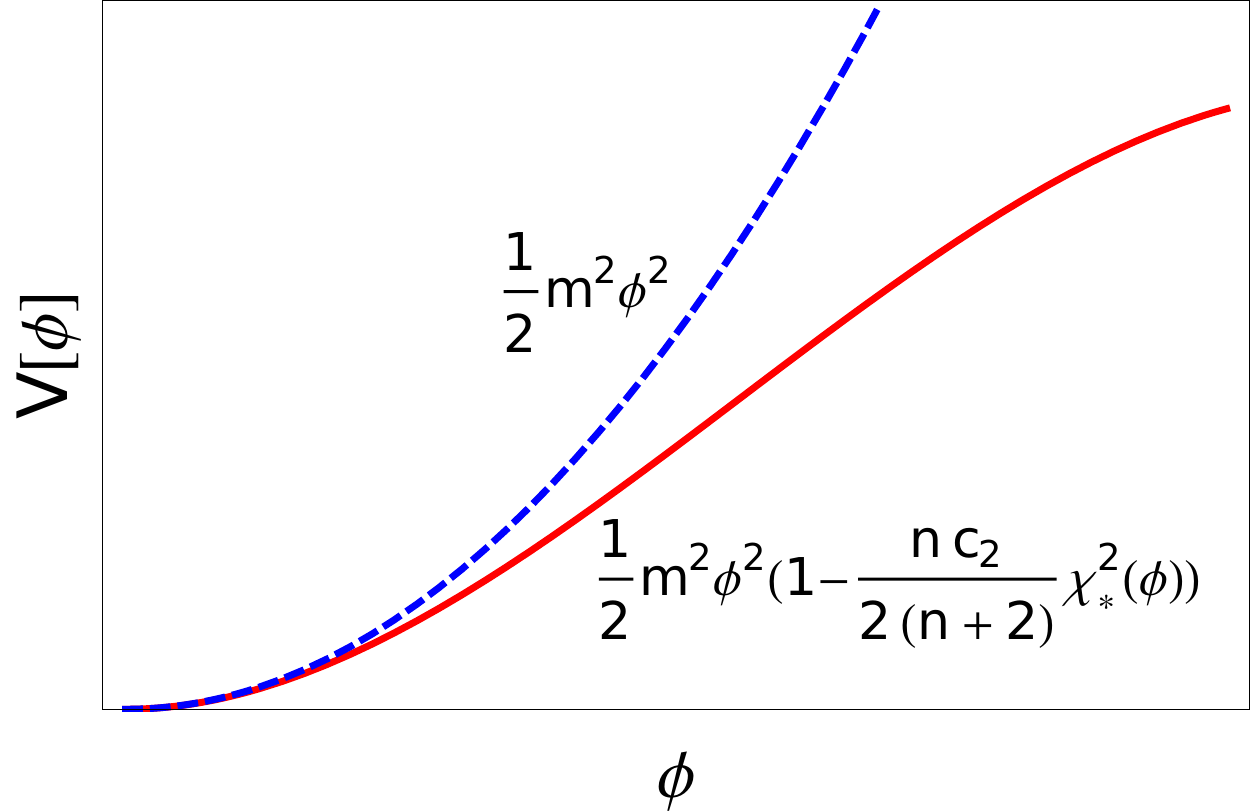}
 \end{center}
\caption{\small\sl Schematic picture of the effective inflaton potential (solid line).
The potential is suppressed than the quadratic potential (dashed line) for a large $\phi$.
}
\label{fig:potential1}
\end{figure}

When $\chi$ is fixed to $\chi_*(\phi)$, the inflaton potential receives back-reaction,
leading to an effective inflaton potential,
\begin{eqnarray}
V(\phi) \simeq \frac{1}{2} m^2 \phi^2 \left(1 -\frac{(n-1)c_2}{n}\chi_*^2(\phi) + \cdots\right)\ .
\end{eqnarray}
Notably, the back-reaction suppresses the inflaton potential at a large inflaton field value. 
In Fig.\,\ref{fig:potential1}, we show a schematic picture of the effective inflaton potential.
There, the inflaton potential is effectively suppressed for a large inflaton field value,
which becomes significant for $\chi_*(\phi) = {\cal O}(0.1)$--${\cal O}(1)$.
As a result, the prediction on $r$, for example, becomes smaller than the one without the back-reacion.
This effect certainly provides a better fit to the observations.

\vspace{.5cm}
The above discussion can be extended to more generic cases,
\begin{eqnarray}
V=\frac{1}{2} m^2 \phi^2 \left(1 - f(\chi)\right) + g(\chi)\ ,
\label{eq:V3}
\end{eqnarray}
where $f(\chi)$ and $g(\chi)$ are some functions of $\chi$.
Due to the softly broken shift symmetry, there is no other terms which couple $\phi$ and $\chi$.%
\footnote{Here, we have neglected terms of ${\cal O}(m^4\phi^4)$. 
We have also assumed a $Z_2$ symmetry under which $\phi$ and $\chi$ are odd, for simplicity.}
As before, we require that the coefficient of the quadratic term of $\chi$ in $f(\chi)$,
is positive and of $O(1)$.
As for a function $g(\chi)$, we require 
\vspace{-3pt}
\begin{enumerate}
\setlength{\itemsep}{-12pt} 
\item $g(\chi)$ is monotonically increasing for $\chi \lesssim {\cal} O(1)$,\\
\item  $g(\chi)$ is shallow so that $g(1) \ll {\cal O}(1)$. 
\end{enumerate}
\vspace{-3pt}
The first condition ensures that  $\chi$ smoothly goes back to a small field value
after inflation.
The second one  allows $\chi$ to have expectation value of ${\cal O}(0.1)$--${\cal O}(1)$ for $\phi \gtrsim 1$.%
\footnote{As we discuss later,  the shallowness of $g(\chi)$ can be easily achieved in supersymmetric models.}

Under these assumptions, $\chi$ is again expelled to $\chi_*(\phi)$ which is determined by,
\begin{eqnarray}
\frac{1}{2}m^2\phi^2f'(\chi_*) = g'(\chi_*)\ .
\end{eqnarray}
Due to the shallowness of  $g(\chi)$,  $\chi_*(\phi)$ is of ${\cal O}(0.1)$--${\cal O}(1)$ 
during inflation,  and hence, the inflaton potential is suppressed due to back-reaction.
It should be emphasized that the suppression is a quite generic feature of this scenario,
since $\chi$ moves to minimize the potential energy for a given value of $\phi$.
We call this mechanism spontaneous suppression of the inflaton potential.

\vspace{.5cm}
Let us pause here and discuss  how  spontaneous suppression affects the initial condition problem.
As we have mentioned, the virtue of the chaotic inflation model is that it satisfies the slow-roll conditions 
even at the Planckian time, i.e. $V\sim M_{\rm PL}^4$ ($\phi \sim 1/m$). 
For such a large $\phi$, one might worry that  $\chi_*(\phi)$ is also 
much larger than $1$ (see Eq.\,(\ref{eq:chistar})).
In reality, however, the higher dimensional terms of $f(\chi)$ become more important
for a larger $\phi$, and $\chi_*(\phi)$ converges to a solution  of $f'(\chi_*) = 0$
and  becomes insensitive to $\phi$. 
With no small parameters in $f(\chi)$, $\chi_*(\infty)$ is naturally expected to be of ${\cal O}(1)$,
and hence, $\chi_*(\phi)$ is of ${\cal O}(1)$ even for $\phi \sim 1/m$.

\begin{figure}[tb]
 \begin{center}
  \includegraphics[width=0.7\linewidth]{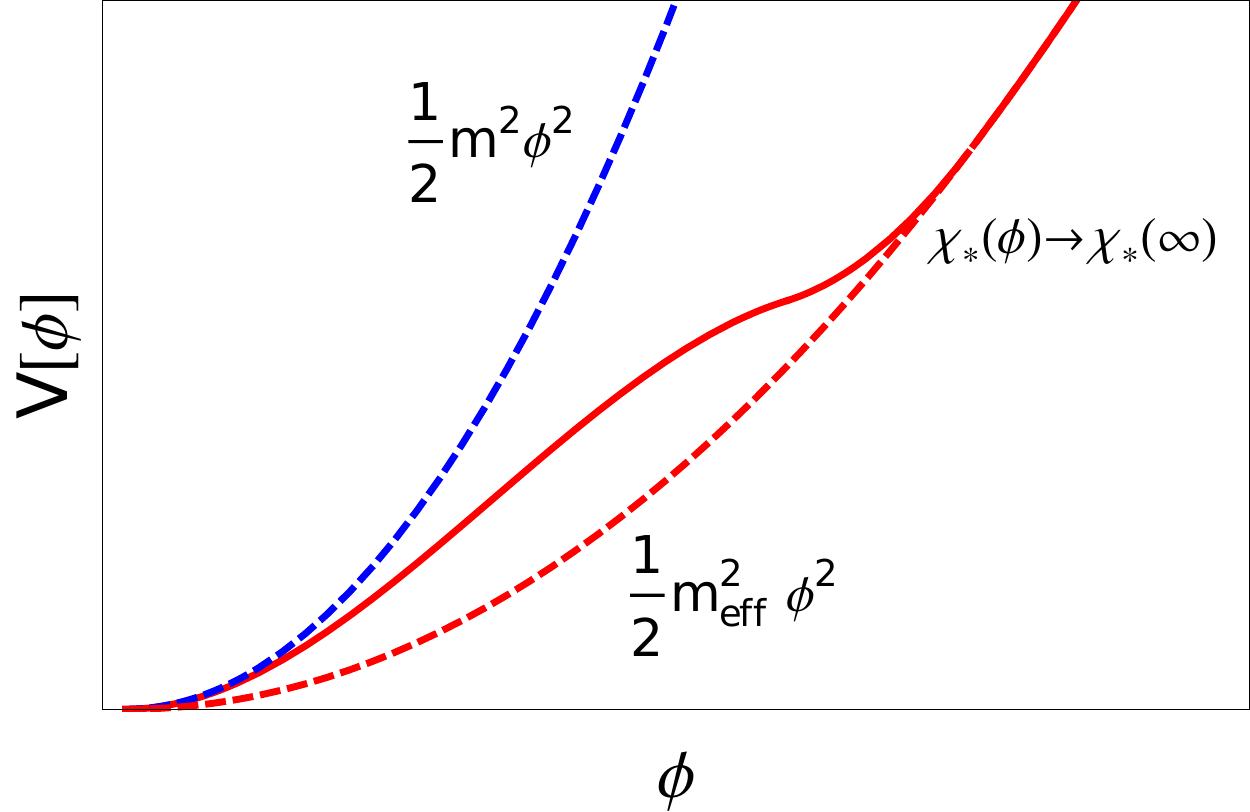}
 \end{center}
\caption{\small\sl Schematic picture of the effective inflaton potential (solid line).
The potential is suppressed than the quadratic potential (upper solid line) for a large inflaton field value
where $\chi_*(\phi) = {\cal O}(0.1)$--${\cal O}(1)$ is achieved.
For a larger inflaton field value, $\chi_*(\phi)$ converges to $\chi_*(\infty)$,
and the inflaton potential becomes quadratic with an effective squared mass, 
$m_{\rm eff}^2 = m^2 (1-f(\chi_*(\infty)))$ (lower solid line).
}
\label{fig:potential2}
\end{figure}

In Fig.\,\ref{fig:potential2}, we show a schematic picture of the effective potential.
As in Fig.\,\ref{fig:potential1},  the inflaton potential is suppressed for a large inflaton field value
where $\chi_*(\phi) = {\cal O}(0.1)$--${\cal O}(1)$ is achieved.
For a much larger $\phi$, $\chi_*(\phi)$ converges to $\chi_*(\infty)$,
and the inflaton potential becomes again quadratic with a slightly smaller mass than $m$.%
\footnote{It should be also noted that the physical mass of the inflaton around its origin 
becomes slightly larger than the one given in Eq.\,(\ref{eq:mass}) for a given magnitude of the curvature perturbation.
This is advantageous to achieve a higher reheating temperature after inflation.}

Several comments are in order.
In Fig.\,\ref{fig:potential2}, we have assumed that $1-f(\chi)$ is positive definite for $\chi<\chi_*(\infty)$.
In particular, if $1-f(\chi_*(\infty)) <0$, the effective inflaton potential is not monotonically increasing 
for a large field value, which screws up the whole picture.
Instead, by assuming $1-f(\chi) >0$ for $\chi<\chi_*(\infty)$, the inflaton potential is always
increasing monotonically, so that the inflaton field smoothly slides down the potential.

To solve the tension between the observations and the original predictions 
of the chaotic inflation model, it is required that
\begin{eqnarray}
\label{eq:conditions}
 {\cal O}(0.1) < \chi_*({\phi_{N_e}}) <  \chi_*(\infty)\ ,~~(N_e \simeq 50-60)\  .
\end{eqnarray}
The first inequality is necessary in order to change the prediction by  ${\cal O}(10)$\%
from the original prediction in Eq.~(\ref{eq:nsr}).
The second inequality is necessary so that the inflaton potential deviates from
a quadratic one at the time of the horizon exit of the pivot scale.
Otherwise, the effective inflaton potential is nothing but the quadratic potential with
a slightly lower mass at that moment, which does not change the predictions on $n_s$ and $r$.%
\footnote{Even if $\chi_*({\phi_{N_e\simeq 50-60}})=\chi_*(\infty)$,
the back-reaction of $\chi$ may leave visible effect on perturbations
if $\chi_*(\phi)$ becomes smaller than $\chi_*(\infty)$ at the horizon exit of smaller scales.
}
In the case of the simplest example give in Eq.\,(\ref{eq:V2}), for example, 
these conditions roughly amount to
\begin{eqnarray}
{\cal O}(0.1)<  \left(\frac{4c_2N_e}{\lambda}m^2\right)^{1/(2n-2)} < {\cal O}(1)\ ,
\end{eqnarray}
which requires $\lambda \sim O(N_e m^2) = O(10^{-9})$.

\vspace{.5cm}
Now, let us demonstrate how spontaneous suppression changes the predictions on $n_s$ and $r$.
In Fig.\,\ref{fig:nsr}, we show the predictions for,
\begin{eqnarray}
f(\chi) &=& c_2 \chi^2 + c_4 \chi^4\ , \\
g(\chi) &=& \frac{\lambda}{4} \chi^4\ .
\end{eqnarray}
In the figure, each line shows the prediction when we change the value of $\lambda$ .
(The parameters,  $c_2$, $c_4$ and $N_e$ are fixed as indicated.)
The five-points-stars denote the predictions for $\lambda \to \infty$. 
There, $\chi$ is not expelled  from its origin, i.e. $\chi_*(\phi_{N_e}) = 0$,
and hence, the predictions coincide with the ones without spontaneous suppression.
We stopped each line when $\chi_*(\phi_{N_e})$ is $\chi_*(\phi_{N_e}) = 2$
to remind the conditions in Eq.\,(\ref{eq:conditions}), 
which roughly correspond to $\lambda \sim 10m^2$. 
The figure shows that the prediction of the minimal chaotic inflation model
is consistent with the observations due to spontaneous suppression.

In our numerical analysis, $\chi$ is completely fixed at  $\chi_*(\phi)$. 
By remembering that the induced mass is not much larger than the Hubble scale, this treatment 
is not very precise. The fluctuation of $\chi$ from $\chi_*(\phi)$ could 
also have visible effect on cosmic perturbations.
We do not pursue this possibility further in this paper, since our main purpose 
is to demonstrate the importance of the back-reaction of $\chi$ to the inflaton dynamics.

\vspace{.5cm}
So far, we have not addressed the naturalness of the shallowness of $g(\chi)$.
To rationalize the shallowness, supersymmetry is the most versatile possibility.
In fact, our arguments can be easily extended to the minimal chaotic inflation
model in supergravity.
The minimal chaotic inflation model can be embedded into  supergravity 
by considering two chiral supermultiplets $\Phi$ and $X$ with the following K\"ahler and the super 
potentials~\cite{Kawasaki:2000yn},
\begin{eqnarray}
K &=& K((\Phi+ \Phi^\dagger)^2,X^\dagger X)  \\
   &=& \frac{1}{2}(\Phi+ \Phi^\dagger)^2   + X^\dagger X + \cdots, \nonumber \\
 W&=&m X \Phi\ .
  \end{eqnarray}
Here, $m$ again denotes a spurion field of the breaking of the shift symmetry, $\Phi \to \Phi + i c $
with $c$ being a real parameter.%
\footnote{Along with softly broken shift symmetry, we also assume a $Z_2$ symmetry and an $R$-symmetry.}
The ellipses denote higher dimensional terms.
Due to the softly broken shift symmetry, the imaginary scalar component of $\Phi \ni i \phi/\sqrt{2}$ obtains 
a quadratic potential in Eq.\,(\ref{eq:V1}) and it plays a role of the inflaton.
The chiral field $X$ is often called a stabilizer, which itself is fixed at its origin during inflation due to
a positive Hubble induced mass.

Now let us introduce a chiral superfield $C$ whose scalar component plays a role of $\chi$ in the above discussion.%
\footnote{Hereafter, $\chi$ denotes a complex scalar field.}
In supersymmetric theory, $\chi$ obtains a negative Hubble induced mass via the  coupling to the stabilizer 
in the K\"ahler potential,
\begin{eqnarray}
K &=& X^\dagger X (1+\tilde f(C^\dagger C))\ , 
\end{eqnarray}
where $\tilde f(C^\dagger C)$ denotes a function of $C^\dagger C$.
The scalar potential of $\chi$  is, on the other hand, obtained from a superpotential,
\begin{eqnarray}
W = \k Y C^n \ ,
\end{eqnarray}
which leads to
\begin{eqnarray}
g(\chi)  = |\k|^2 |\chi|^{2n} \ .
\end{eqnarray}
Here $Y$ denotes another chiral superfield, and $\k$ the coupling constant.
As a virtue of a superpotential term, there is no wonder to have a very suppressed $\k$.
Thus, in supersymmetric model, the shallow scalar potential $g(\chi)$ can be 
easily rationalized.
It should be also emphasized that flat directions are ubiquitous 
in supersymmetric theory (such as $D$ and $F$ flat directions
in the supersymmetric standard model).
Therefore, the scenario with spontaneous suppression goes particularly well  with supersymmetry.%
\footnote{In fact, a flat direction in the supersymmetric standard model which takes
 a large field value play a central role in the Affleck-Dine Baryogenesis~\cite{Affleck:1984fy,Dine:1995kz}.}

\vspace{.5cm}
In this letter, we pointed out that the prediction of minimal chaotic inflation models is
affected by the existence of some scalar fields when they obtain large field values of 
${\cal O}(0.1)$--${\cal O}(1)$ during inflation due to negative Hubble induced  masses.
As we demonstrated, the prediction of the quadratic chaotic inflation model is 
consistent with observational constraints on $n_s$ and $r$
when we take spontaneous suppression into account.
Let us emphasized that scalar fields with negative Hubble induced masses are rather common 
(in particular in supersymmetry models), and hence, it is important to keep this effect in our mind.

It should be also noted that spontaneous suppression of the inflaton potential occurs
in other large field inflation models such as the natural inflation models.
Therefore, the predictions of those models are also altered by spontaneous suppression.

Finally, we comment on a possible connection between spontaneous suppression and the standard model.
To date, the Higgs boson is the only known scalar field to exist apart from the inflaton.
Interestingly, the Higgs quartic coupling seems almost vanishing at around the 
Planck scale within the uncertainties of standard model parameters
(see e.g.~\cite{Holthausen:2011aa,EliasMiro:2011aa,Degrassi:2012ry,Buttazzo:2013uya}).
This shows that the Higgs boson is a candidate of $\chi$.
This possibility can be partially tested by future precise measurements 
of the Higgs mass parameters, the top Yukawa coupling and the strong coupling constants.


\vspace{.7 cm}

%
\headline{Acknowledgements}\\
This work is supported by Grants-in-Aid for Scientific Research from the Ministry of Education, Culture, Sports, Science, and Technology (MEXT), Japan,
No. 24740151 and No. 25105011 (M.~I.),
No. 25400248 (M.~K.)
as well as No. 26104009 (T.~T.~Y.); 
Grant-in-Aid No. 26287039 (M.~I. and T.~T.~Y.) from the Japan Society for the Promotion of Science (JSPS); and by the World Premier International Research Center Initiative (WPI), MEXT, Japan (M.~I., M.~K. and T.~T.~Y.).
K.H. is supported in part by a JSPS Research Fellowship for Young Scientists.
%


\begin{thebibliography}{99}

\bibitem{Guth:1980zm} 
  A.~H.~Guth,
  Phys.\ Rev.\ D {\bf 23}, 347 (1981).

\bibitem{Kazanas:1980tx} 
  D.~Kazanas,
  Astrophys.\ J.\  {\bf 241}, L59 (1980).

\bibitem{Linde:1981mu} 
  A.~D.~Linde,
  Phys.\ Lett.\ B {\bf 108}, 389 (1982).

\bibitem{Albrecht:1982wi} 
  A.~Albrecht and P.~J.~Steinhardt,
  Phys.\ Rev.\ Lett.\  {\bf 48}, 1220 (1982).

\bibitem{Starobinsky:1980te} 
  A.~A.~Starobinsky,
  Phys.\ Lett.\ B {\bf 91}, 99 (1980).

\bibitem{Mukhanov:1981xt} 
  V.~F.~Mukhanov and G.~V.~Chibisov,
  JETP Lett.\  {\bf 33}, 532 (1981)
  [Pisma Zh.\ Eksp.\ Teor.\ Fiz.\  {\bf 33}, 549 (1981)];
%
\bibitem{Hawking:1982cz} 
  S.~W.~Hawking,
  Phys.\ Lett.\ B {\bf 115}, 295 (1982).
\bibitem{Starobinsky:1982ee} 
  A.~A.~Starobinsky,
  Phys.\ Lett.\ B {\bf 117}, 175 (1982).
\bibitem{Guth:1982ec} 
  A.~H.~Guth and S.~Y.~Pi,
  Phys.\ Rev.\ Lett.\  {\bf 49}, 1110 (1982).
\bibitem{Bardeen:1983qw} 
  J.~M.~Bardeen, P.~J.~Steinhardt and M.~S.~Turner,
  Phys.\ Rev.\ D {\bf 28}, 679 (1983).

\bibitem{Linde:1983gd} 
  A.~D.~Linde,
  Phys.\ Lett.\ B {\bf 129}, 177 (1983).

\bibitem{Linde:2005ht} 
  A.~D.~Linde,
  Contemp.\ Concepts Phys.\  {\bf 5}, 1 (1990)
  [hep-th/0503203].
  

\bibitem{Kawasaki:2000yn} 
  M.~Kawasaki, M.~Yamaguchi and T.~Yanagida,
  Phys.\ Rev.\ Lett.\  {\bf 85}, 3572 (2000)
  [hep-ph/0004243].
\bibitem{Ade:2015lrj} 
  P.~A.~R.~Ade {\it et al.}  [Planck Collaboration],
  arXiv:1502.02114 [astro-ph.CO].


\bibitem{Kallosh:2010ug}
  R.~Kallosh and A.~Linde,
  JCAP {\bf 1011}, 011 (2010)
  [arXiv:1008.3375 [hep-th]].


\bibitem{Li:2013nfa} 
  T.~Li, Z.~Li and D.~V.~Nanopoulos,
  JCAP {\bf 1402}, 028 (2014)
  [arXiv:1311.6770 [hep-ph]].
  
  

\bibitem{Harigaya:2014qza} 
  K.~Harigaya and T.~T.~Yanagida,
  Phys.\ Lett.\ B {\bf 734}, 13 (2014)
  [arXiv:1403.4729 [hep-ph]];
  K.~Harigaya, M.~Kawasaki and T.~T.~Yanagida,
  Phys.\ Lett.\ B {\bf 741}, 267 (2015)
  [arXiv:1410.7163 [hep-ph]].

\bibitem{Silverstein:2008sg} 
 E.~Silverstein and A.~Westphal,
 Phys.\ Rev.\ D {\bf 78}, 106003 (2008),
 0803.3085 [hep-th];
 L.~McAllister, E.~Silverstein and A.~Westphal,
 Phys.\ Rev.\ D {\bf 82}, 046003 (2010),\newline
 0808.0706 [hep-th].

\bibitem{Takahashi:2010ky} 
 F.~Takahashi,
 Phys.\ Lett.\ B {\bf 693}, 140 (2010),
 1006.2801 [hep-ph].

\bibitem{Harigaya:2012pg} 
 K.~Harigaya, M.~Ibe, K.~Schmitz and T.~T.~Yanagida,
 Phys.\ Lett.\ B {\bf 720}, 125 (2013)
 [arXiv:1211.6241 [hep-ph]];
 Phys.\ Lett.\ B {\bf 733}, 283 (2014)
 [arXiv:1403.4536 [hep-ph]];
 Phys.\ Rev.\ D {\bf 90}, no. 12, 123524 (2014)
 [arXiv:1407.3084 [hep-ph]].

\bibitem{Destri:2007pv} 
  C.~Destri, H.~J.~de Vega and N.~G.~Sanchez,
  Phys.\ Rev.\ D {\bf 77}, 043509 (2008)
  [astro-ph/0703417].

\bibitem{Croon:2013ana} 
  D.~Croon, J.~Ellis and N.~E.~Mavromatos,
  Phys.\ Lett.\ B {\bf 724}, 165 (2013)
  [arXiv:1303.6253 [astro-ph.CO]].

\bibitem{Nakayama:2013jka} 
  K.~Nakayama, F.~Takahashi and T.~T.~Yanagida,
  Phys.\ Lett.\ B {\bf 725}, 111 (2013)
  [arXiv:1303.7315 [hep-ph]];
  JCAP {\bf 1308}, 038 (2013)
  [arXiv:1305.5099 [hep-ph]].

\bibitem{Affleck:1984fy} 
  I.~Affleck and M.~Dine,
  Nucl.\ Phys.\ B {\bf 249}, 361 (1985).



\bibitem{Dine:1995kz} 
  M.~Dine, L.~Randall and S.~D.~Thomas,
  Nucl.\ Phys.\ B {\bf 458}, 291 (1996)
  [hep-ph/9507453].

\bibitem{Holthausen:2011aa} 
  M.~Holthausen, K.~S.~Lim and M.~Lindner,
  JHEP {\bf 1202}, 037 (2012)
  [arXiv:1112.2415 [hep-ph]].
\bibitem{EliasMiro:2011aa} 
  J.~Elias-Miro, J.~R.~Espinosa, G.~F.~Giudice, G.~Isidori, A.~Riotto and A.~Strumia,
  Phys.\ Lett.\ B {\bf 709}, 222 (2012)
  [arXiv:1112.3022 [hep-ph]].
\bibitem{Degrassi:2012ry} 
  G.~Degrassi, S.~Di Vita, J.~Elias-Miro, J.~R.~Espinosa, G.~F.~Giudice, G.~Isidori and A.~Strumia,
  JHEP {\bf 1208}, 098 (2012)
  [arXiv:1205.6497 [hep-ph]].
\bibitem{Buttazzo:2013uya} 
  D.~Buttazzo, G.~Degrassi, P.~P.~Giardino, G.~F.~Giudice, F.~Sala, A.~Salvio and A.~Strumia,
  JHEP {\bf 1312}, 089 (2013)
  [arXiv:1307.3536 [hep-ph]].

\end{thebibliography}
\end{document}